\documentclass[preprint,aps,showpacs]{revtex4}
\usepackage{graphicx}

\usepackage{epsfig}
\usepackage{subfigure}
\input{epsf.sty}

\newcommand{\beq}{\begin{equation}}
\newcommand{\eeq}{\end{equation}}
\newcommand{\ba}{\begin{array}}
\newcommand{\ea}{\end{array}}
\newcommand{\beqa}{\begin{eqnarray}}
\newcommand{\eeqa}{\end{eqnarray}}
\newcommand{\bd}[1]{ \mbox{\boldmath $#1$}  }

\newcommand{\lam}{\lambda}

\newcommand{\APNY}[1]{Ann. Phys. (N.Y.) {\bf {#1}}}

\newcommand{\JPG}[1]{J. Phys. {\bf G{#1}}}
\newcommand{\NPA}[1]{Nucl. Phys. {\bf A{#1}}}
\newcommand{\PLB}[1]{Phys. Lett. {\bf B{#1}}}

\newcommand{\PRC}[1]{Phys. Rev. C {\bf {#1}}}
\newcommand{\PRL}[1]{Phys. Rev. Lett. {\bf {#1}}}
\newcommand{\PR}[1]{Phys. Rep. {\bf {#1}}}

\newcommand{\RMP}[1]{Rev. Mod. Phys. {\bf {#1}}}

\begin{document}

\title{Signature of Shallow Potentials in Deep \\
Sub-barrier Fusion Reactions}
\author{\c S.~Mi\c sicu\footnote{on leave of absence from
National Institute for Nuclear Physics, Bucharest, P.O.Box MG6,
Romania.} and H.~Esbensen}
\affiliation{Physics Division, Argonne National Laboratory, Argonne, Illinois 60439, USA}
\date{\today}

\begin{abstract}
We extend a recent study that explained the steep falloff in the
fusion cross section at energies far below the Coulomb barrier for the
symmetric dinuclear system  $^{64}$Ni+$^{64}$Ni to another symmetric system, 
$^{58}$Ni+$^{58}$Ni, and the asymmetric system $^{64}$Ni+$^{100}$Mo.
In this scheme the very sensitive dependence of the internal part of the nuclear
potential on the nuclear equation of state determines a reduction of the classically
allowed region for overlapping configurations and consequently a decrease in the 
fusion cross sections at bombarding energies far below the barrier.
Within the coupled-channels method, including couplings to the low-lying 2$^+$
and 3$^-$ states in both target and projectile as well as mutual and two-phonon 
excitations of these states, we calculate and compare with the experimental fusion 
cross sections, $S$-factors, and logarithmic derivatives for the above mentioned 
systems and find good agreement with the data even at the lowest energies.
We predict, in particular, a distinct double peaking in the $S$-factor for the far 
subbarrier fusion of $^{58}$Ni+$^{58}$Ni which should be tested experimentally. 
\end{abstract}
\maketitle

\section{Introduction}

Very recently we proposed a mechanism that can explain the
unexpected hindrance or steep falloff of fusion cross sections 
which has been observed
at bombarding energies far below the Coulomb barrier \cite{misesb06a}. 
Although measurements of several medium-heavy nuclei performed in the past two 
decades already provided some indications of a steep decrease of the excitation 
functions at the lowest bombarding energies, the credit of disclosing and confirming 
this unexpected trend for new fusing systems has to be given to C. L. Jiang et al. 
\cite{jiang02,jiang04a,jiang04b,jiang05,jiang06}. 
Among the most conspicuous cases reported in the past are $^{58}$Ni+$^{58}$Ni
\cite{beck82}, where the departure from the expected behavior takes places already 
at cross sections $\sim$ 0.1 mb, and $^{90}$Zr+$^{89}$Y, $^{90}$Zr+$^{92}$Zr
as reported in \cite{kel86}.

The new fusion data reported by Jiang et al. are even more spectacular 
because the reported cross sections are measured down to 10$~$nb :
$^{60}$Ni+ $^{89}$Y \cite{jiang02} ($\sigma_f\geq 100~$nb), 
$^{64}$Ni+ $^{64}$Ni \cite{jiang04b} ($\sigma_f\geq 10~$nb),  
$^{64}$Ni+$^{100}$Mo \cite{jiang05} ($\sigma_f > 10~$nb).
The hindrance of fusion was first reported as a suppression of the 
measured low-energy fusion cross sections with respect to model 
calculations \cite{jiang02}.
It was later on characterized by the energy $E_s$ where the $S$-factor 
for fusion develops a maximum at low energy \cite{jiang04a}.
Following the publication of these findings a challenge was launched, 
especially on the theoretical side of the heavy-ion fusion community.  
At the end of 2005 the underlying physical cause of this apparently new 
phenomenon was still unknown according to the authors of Ref.~\cite{jiang06}.
Some authors have even advocated the hypothesis that the standard theoretical 
approach to treat capture reactions, i.~e. the coupled-channels (C.~C.) method 
is unable to explain the steep falloff of the cross sections.
We shall try to convince the reader throughout this paper that the C.~C. 
method is the right tool to investigate capture, even at very low energy.

The nuclear potentials that are employed in C.~C. calculations are commonly
parametrized as a Woods-Saxon well.  Among the issues related to the deep 
sub-barrier fusion
was the large diffuseness $a$ of the ion-ion potential that was needed to fit 
high-precision fusion data. Hagino et al. \cite{hagrodas03}, hinted that a 
phenomenological nuclear potential with a larger diffuseness leads to a better 
agreement to the data. Values up to $a=1.3$ fm were conjectured instead of the usual 
$a=0.65$ fm for the system $^{58}$Ni+$^{58}$Ni. 
In \cite{jiang04a} it was remarked that since the low-energy fusion becomes sensitive 
to the nuclear potential inside the barrier, this part of the interaction may not be 
accurately modeled by the conventional Woods-Saxon parametrization. It was pointed out 
that by doubling the diffuseness of the inner part of the potential, the agreement 
with the data improves for the colliding system $^{60}$Ni+$^{89}$Y \cite{jiang04a}. 

Following the same idea, the systematic failure of the Woods-Saxon nuclear potential 
to describe fusion was analyzed in \cite{new04} and it was concluded that the origin 
of it should be sought in the diffuseness parameter $a$. In order to fit the data 
at energies above the Coulomb barrier, the diffuseness must be increased with 
increasing $Z_1Z_2$. The fusion data compiled by these authors indicate a correlation 
with the neutron richness of the projectile 
and target nuclei in the sense that the neutron rich nuclei tend to require larger 
values of $a$. 

The authors of \cite{dass03} pointed out that potentials such as the 
Aky\"uz-Winther (AW) \cite{browin91,akywin} are providing reliable barriers 
but they cannot reproduce the data far below the 
barrier, a fact which made them suggest that the 
ion-ion potential should have another form in the inner part of the barrier. 
Following a sequence of simple but clear arguments they pointed out that the 
exponential falloff in the tunneling probability is related to the disappearance 
of the classically allowed region below a certain energy. 
If this is true, it would mean that we are confronted with the existence of a 
shallow pocket of the potential inside the barrier. 

In Ref.~\cite{seif04} a surprisingly good description of the data for 
$^{58}$Ni+$^{58}$Ni,  $^{64}$Ni+$^{64}$Ni and  $^{60}$Ni+$^{89}$Y
was claimed. However, it is difficult to judge the significance of the 
results because of the limited number of excitations that were included 
in the C.~C. calculations. For example, the only excitation considered 
in the nickel isotopes is the one-phonon $2^+$ excitation,
but it is well known that couplings to the $3^-$ state and higher-order 
couplings to two-phonon states are tremendously enhancing the 
heavy-ion fusion cross section \cite{esb-land87,hag97}.

Other works have attempted to account for the complicated problem of channel 
coupling by means of a polarization potential \cite{lin04}. 
The imaginary potential parametrizes in this case the excitation of 
other degrees of freedom that influence the fusion process. 
According to \cite{lin04}, the imaginary potential shows a rapid cutoff 
at energies far below the Coulomb barrier for the cases $^{58}$Ni+ $^{58}$Ni,
$^{58}$Ni+ $^{64}$Ni and $^{64}$Ni+ $^{64}$Ni because of a threshold below 
which the C.~C. effects are ceasing to exist. 

Before ending the list of hypotheses advanced to explain the fusion
hindrance phenomenon we quote the result reported in Ref.~\cite{gir04}, 
which overrules the possibility of explaining the depletion of fusion
rates at extreme sub-barrier energies from dissipative tunneling, as  
results from other quantum open system approaches, such as that of Caldeira and
Leggett \cite{cal-leg82}.

Very recently \cite{misesb06a} we proposed an explanation of the hindrance 
observed in the sub-barrier fusion of  $^{64}$Ni+$^{64}$Ni, which was based 
on the same standard coupled-channels formalism as before \cite{jiang04a} 
but with amendments that concern the potential. 
Essential in getting a good description of the data was to take into account 
the saturation of nuclear matter and to use realistic neutron and proton 
distributions of the reacting nuclei. 
These two  ingredients are naturally incorporated in a potential calculated 
via the double-folding method with tested effective nucleon-nucleon forces 
and with realistic charge and nuclear densities, a fact which is often 
overlooked or only indirectly included in the Woods-Saxon parametrization.
We intend in this paper to show that by properly addressing these
issues, light can be shed on the extreme sub-barrier fusion data.
Before we do that we summarize the methods that have been used to 
analyze the low-energy behavior of heavy-ion fusion cross sections.

\section{Representations of low-energy cross sections} 

In an attempt to get a diagnosis for the various cases where the hindrance in 
sub-barrier fusion occurs, the authors of Ref.~\cite{jiang04a} proposed the 
use of two representations.
The first one is the astrophysical $S$-factor \cite{burb57}, 
\beq
S = E\sigma(E) \ \exp(2\pi\eta),
\eeq
where $E$ is the center-of-mass energy, $\eta=Z_1Z_2e^2/(\hbar v_{\rm rel})$ is the 
Sommerfeld parameter and $v_{\rm rel}$ is the relative velocity of the fragments.
The experimental value of $S$ increases with decreasing bombarding energy
and has the tendency to develop a maximum for the systems of interest. 
The necessity to resort to this quantity comes from the fact that the reaction
cross section varies by many orders of magnitude below the Coulomb barrier
(7 orders of magnitude for $^{64}$Ni+ $^{64}$Ni). The $S$-factor has been used in 
the past to unravel typical molecular resonant structures in the excitation 
function of systems like $^{12}$C+$^{12}$C since it removes the dominating influence 
of the Coulomb and centrifugal barrier transmission factors that mask these
structures in the cross section \cite{erb85}. The series of narrow and prominent
resonances was associated with quasi-bound, long-lived states of the $^{24}$Mg
nucleus.  

Very recently the fusion cross section for the $^{12}$C+$^{12}$C system
has been measured down to very low energies \cite{barron}. The data show 
a rise in the $S$-factor at the lowest energies, which might indicate the 
existence of a broad resonance in the entrance channel, possibly related 
to an intermediate state in the compound nucleus \cite{barron}. 
Similar broad resonances in the $S$-factor at energies below the barrier
have been also inferred from the $^{12}$C+ $^{16}$O total cross sections 
\cite{pat71}. Thus, the $S$-factor is a quantity that magnifies structures 
in the excitation function at energies below the barrier,
and it is also an instrument for exploring the inner part 
of the barrier in low-energy, heavy-ion fusion reactions.

A second representation proposed in Ref. \cite{jiang04a} is the logarithmic 
derivative,
\beq
L(E)=\frac{d[\ln(E\sigma)]}{dE}=\frac{1}{E\sigma}\frac{d}{d~E}(E\sigma).
\eeq
The point where the experimental $L(E)$ intersects the logarithmic derivative 
obtained from an $s$-wave transmission across a pure point-charge Coulomb 
potential (constant $S$-factor), given by $L_0(E)=\frac{\pi\eta}{E}$, 
coincides with the maximum in the $S$-factor invoked earlier. 
Extracting the energy $E_s$, where this intersection occurs,
it was found that the corresponding logarithmic derivative $L_s$ is nearly 
the same for stiff heavy-ion systems, with an average value of 2.34 
MeV$^{-1}$. This implies that $E_s$ scales with the charge and mass numbers 
of the reacting nuclei according to the empirical law \cite{jiang04a}
\beq
E_s=0.355\left [ Z_1Z_2\sqrt{\frac{A_1A_2}{A_1+A_2}}\right ]^{2/3}.
\eeq  

At this point it is useful to recall that the inability of previous calculations 
to reproduce the low-energy data points of the measured fusion cross sections is 
most clearly seen from the inspection of the logarithmic derivative $L(E)$.
Thus for energies below a certain threshold, the experimental values of $L(E)$
increase steeply with decreasing energy, whereas the theoretical curve increases 
with a much smaller slope on which a resonant structure is superposed.    

\section{Simulations of a repulsive core}

The occurrence of resonant structures in collisions of light nuclei,
the best known example being the sharp peaks
in the bombarding energy dependence of the gamma radiation yields emerging 
in the  $^{12}$C+$^{12}$C scattering found by Bromley et al. \cite{brom60},
were found to resemble states in a molecular potential well.
Following the suggestion that these ''quasi-molecular'' states may represent 
doorway states to fusion, the resonant behavior found its explanation by introducing 
the concept of the {\em double-resonance mechanism} \cite{scheid70}. 
In this scenario the indirect population of quasi-molecular states can occur 
in light-ion scattering according to the following sequence of events: 
a) Surpassing the potential barrier at an initial energy $E_i$ and losing 
the energy $E^*$ by inelastic excitations of low-energy 
levels of one or both of the ions, and 
b) Resonant decay into the potential pocket, where the colliding nuclei are trapped 
and a quasi-molecule is formed, if their relative energy $E_i-E^*$ coincides with 
the energy of a quasi-bound state.    

Evidence that the  resonant behavior observed in 
$^{12}$C+$^{12}$C is not an isolated phenomenon
was made available for other light systems such as 
$^{12}$C+ $^{16}$O,  $^{12}$C+ $^{13}$C, $^{16}$O+ $^{16}$O, $^{16}$O+ $^{24}$Mg.
It persists even in heavier colliding systems, such as the  $^{24}$Mg+ $^{24}$Mg
\cite{zur84} and $^{28}$Si+ $^{28}$Si \cite{bet81}.
Manifestation of clusterization in connection to quasi-molecular pockets is also 
known for heavy nuclear systems, such as the cluster radioactivity \cite{rose84}, 
or hyperdeformation and clustering in the actinide region \cite{hamilt94,krasz02}. 

To simulate the appearance of shallow pockets several recipes have been 
proposed :
\begin{itemize}
\item A central soft repulsive core added to the conventional
Woods-Saxon potential was used in Refs.~\cite{mbm68} and \cite{mich73} in order 
to fit the reaction cross sections
observed in $^{12}$C+ $^{12}$C,  $^{12}$C+ $^{16}$O and $^{16}$O+ $^{16}$O.
The effect of a repulsive core was modeled by a positive Gaussian potential,
$V_{\rm rep}\exp(-br^2)$, the width $b$ being constrained by the requirement that 
the potential becomes repulsive for $r\leq R_0$, $R_0$ being the radius of the 
Woods-Saxon potential. 
\item Double-folding potentials as introduced by Satchler and Love 
\cite{satlov79} are accurate only in the tail region of the 
nucleus-nucleus potential, where the density distributions are only gently
overlapping and thus the assumption of ''frozen density'' is less questionable. 
However this assumption ignores any readjustment due to the mutual interaction 
of the nuclei or the Pauli exclusion principle for strong overlap. 
To cope with this  problem the observation was made in \cite{seiw84,seiw85} that, 
whereas any theory of heavy-ion potentials is expected to reduce to 
the double-folding model in the limit of large ion-ion separations and vanishing 
density overlap, the compound system resulting from fusing the two ions is accurately 
described by the liquid-drop energy $E_{\rm LDM}$. To interpolate between these two 
extremes the nuclear deformation energy was written as
\beq
E_{\rm def}=E_{\rm LDM}+W_{M3Y+C}(\rho)-\alpha \overline{W}_{M3Y}(\bar\rho),
\eeq 
where $W_{M3Y+C}(\rho)$ is the double folding self-interaction energy with the 
original M3Y parametrization for the nuclear and Coulomb forces with realistic 
values of the proton and neutron matter diffusivities and radii. The  
$\overline{W}_{M3Y}(\bar\rho)$ is a renormalization introduced with the purpose to 
subtract the volume and surface energy contributions in $W$, and is computed with 
sharp-edge densities.  The parameter 
$\alpha$ is adjusted so that $W$ and $\overline{W}$ cancel exactly in the limit of 
complete overlap. This scheme was applied only for collisions of two identical 
spherical nuclei since  the evaluation becomes to cumbersome for deformed nuclei 
approaching at different orientations. 

\item 
Pockets in a double folding potential are arising naturally if an effective 
Skyrme like $NN$ force is used \cite{misprot98}. 
However, this force was not tested systematically for scattering
reactions as is the case for the M3Y force.

\item 
Density-dependent interactions superposed on the original M3Y form have been 
used in recent times to simulate the saturation of nuclear matter for 
$\alpha$+nucleus scattering or elastic scattering of light nuclei 
\cite{brasat97,khoa97}.
An attempt to explain fusion data for a variety of systems 
using these density dependent interactions was 
made very recently by the Canberra group \cite{gont04} and they reached
the conclusion that the double-folding model may not be appropriate
for fusion. 
The pockets resulting from the density dependence of the effective 
$NN$ force are still too deep
to improve the agreement of C.~C. calculations and the experimental data 
at extreme sub-barrier energies.

\item Methods based on the energy density functional are known for a long time 
to be able to predict shallow pockets in the interaction between two nuclei 
\cite{blomal67,scheid68}. A typical feature of this class of potentials is the 
short-ranged  repulsion at distances where a strong overlap of the nuclear
densities takes place. 
In this approach the condition of nuclear matter saturation is achieved by fitting
the free parameters in the energy-density functional using the 
properties of finite nuclei. 
Like in the case of the double-folding recipe, 
the sudden-approximation, i.e. the summation of frozen nuclear densities,
is essential for the occurrence of the core.
Very recently, using this method, in the framework of the extended Thomas-Fermi 
approximation with $\hbar^2$ correction terms in the kinetic
energy density functional, an analytic potential with parameters that are fitted to 
data was proposed \cite{denis02}. However, this potential provides 
barriers that are higher than the experimental ones and they can therefore not 
provide a satisfactory description of the data in the barrier region.  
\item Proximity potentials are well-known examples of nuclear interactions 
producing pockets in the ion-ion potential
(see Ref.~\cite{block77} for the 1977  and \cite{myswy00} for the 2000 versions). 
Unfortunately these potentials provide pockets that are too deep for the systems 
of interests (see for example Fig. 1 of Ref. \cite{misesb06a}).
\item Finally we mention the semi-empirical Aky\"uz-Winther potential
which we will refer to in the following as the AW potential.
It is parametrized as a Woods-Saxon potential \cite{browin91}, so that the 
maximum nuclear force  is consistent with the proximity force for touching 
spheres, and it has an exponential tail that is consistent with the 
M3Y double-folding calculated in Ref. \cite{akywin}. The AW also produces
pockets that are too deep as evidenced by its applications in Refs.
\cite{misesb06a,jiang02,jiang04a,jiang04b,jiang05,jiang06}.
\end{itemize} 

In the present work we are interested in the interaction between two 
intermediate mass nuclei.
For these target-projectile combinations the above approach to the 
density dependent interaction does not necessarily apply because, 
as we shall see, the resulting pockets are still to deep. 
For this reason another approach will be adopted.

\section{The M3Y+repulsion, double-folding Potential}

As a manifestation of the Pauli principle, which prevents the overlapping
of the wave functions of two systems of fermions, we expect that the interaction 
potential between two colliding nuclei will contain a repulsive core.
The actual form of the repulsive core and its strength  depend strongly on 
the extent to which the collision is adiabatic or sudden. 
A further uncertainty with regard to the core parameters is  
the influence of individual characteristics of the considered nuclei, 
including binding energies, shapes, and the nucleon distributions.
The models of ion-ion potentials which provide a repulsive core and which 
we enumerated in the preceding section, lead to quantitatively different 
estimates of the height, radius and diffuseness of the core potential.
Due to these conditions of strong uncertainty we use like in previous works
on fusion \cite{misgre04} a crude recipe to determine the properties of the
repulsive core potential.

\subsection{Calibration of the repulsive core potential}

As noticed earlier, an overlapping region with doubled nucleon density 
is formed once the distance $r$ between the nuclei becomes less than 
$R_1+R_2$, where $R_1$ and $R_2$ are the nuclear radii of the target 
and projectile along the scattering (fission) axis. 
Here and in the following we assume for simplicity that the densities of 
the two nuclei are frozen when estimating the nuclear interaction potential.
When a complete overlap of the two nuclei occurs, the total density is 
therefore roughly twice that of normal matter, $\rho\approx~2\rho_0$,
within the volume of the smaller nucleus.
Consequently, the nuclear equation-of-state (EOS) dictates an increase 
$\Delta V$ in the energy of the compound system. 
In the case of complete overlap (for $r=0$) $\Delta V$ is proportional 
to the increase of the energy per particle of nuclear matter 
$\varepsilon(\rho,\delta)$ considered as a function of the nuclear density, 
$\rho=\rho_n+\rho_p$, and the relative neutron excess, 
$\delta=(\rho_n-\rho_p)/\rho$,
\beq
\Delta V \approx 2 A_p\left [  \varepsilon(2\rho_0,\delta)
-\varepsilon(\rho_0,\delta)\right ].
\label{varpotrep}
\eeq
Here the relative neutron excess $\delta$ is assumed to be approximately 
constant, and $A_p$ is the mass number of the smaller nucleus in the case 
of an asymmetric system.  

The EOS predicted by the Thomas-Fermi model for cold nuclear matter is \cite{myswy98}
\beq
\varepsilon(\rho,\delta)=\varepsilon_F
\left [ A(\delta)\left (\frac{\rho}{\rho_0}\right )^{2/3} 
       + B(\delta)\left (\frac{\rho}{\rho_0}\right )
       + C(\delta)\left (\frac{\rho}{\rho_0}\right )^{5/3}\right ],
\label{nucleos}
\eeq
where $\rho_0=0.161$ fm$^{-3}$ is the saturation
density and $\varepsilon_F$ is the Fermi energy of  normal nuclear matter.
The expressions for the $\delta$-dependence entering Eq.~(\ref{nucleos})
are listed in Ref. \cite{myswy98}.
The definition of the incompressibility of cold nuclear matter is
then related to the curvature of $\varepsilon(\rho,\delta)$ at 
the saturation density,
\beq
K=9\left ( \rho^2\frac{\partial^2 \varepsilon}{\partial\rho^2}
\right )_{\rho=\rho_{0}}.
\eeq

In order to calibrate the strength of the repulsive core potential,
we assume that $\Delta V$ in Eq.~(\ref{varpotrep}) must be identified 
with the nuclear part of the heavy-ion potential, $V_N(r)$, evaluated 
at the coordinate origin $r=0$, 
\beq
\Delta V = V_{N}(0).
\eeq 
Approximating the EOS by a parabolic expansion
around the equilibrium density $\rho_0$ \cite{eisgre87},
\beq
\varepsilon(\rho,\delta)=\varepsilon(\rho_0,\delta)
+\frac{K}{18\rho_0^2}\ (\rho-\rho_0)^2,
\eeq
we obtain the following estimate of the nuclear potential
at $r=0$,
\beq
\label{DVCAL}
V_{N}(0) = \Delta V \approx \frac{A_p}{9} \ K.
\eeq
Similar recipes for introducing a repulsive core can be found in the
literature \cite{ueg93}. They are based on the knowledge of the 
equation of state and on the requirement that the nuclear matter 
density is doubled  for a total overlap of the two reacting nuclei. 

\subsection{Folding Model Potential}
  
We consider two nuclei with one-body  deformed densities $\rho_1$ 
and $\rho_2$, subjected to vibrational fluctuations, and center of 
masses separated by the distance $\bd{r}$.
Then the nuclear potential between these two nuclei can be evaluated
as the double folding integral,
\beq
V_{N}(\bd{r}) = \int d\bd{r}_1 \int d\bd{r}_2 \
\rho_1(\bd{r}_1) \ \rho_2(\bd{r}_2) \ v(\bd{r}_{12}), 
\label{dfold}
\eeq
where $\bd{r}_{12}=\bd{r}+\bd{r_2}-\bd{r}_1$.
The central part of the effective $NN$ interaction $v(\bd{r}_{12})$ contains 
no spin or spin-isospin terms since we analyze fusion reactions of spin 0 
nuclei, where the spin terms are relatively unimportant.
We are then left with a direct part,
\beq
v_{\rm dir}(\bd{r}_{12})=v_{00}(\bd{r}_{12})
+\frac{N_1-Z_1}{A_1}\ \frac{N_2-Z_2}{A_2}\ v_{01}(\bd{r}_{12}),
\label{vdir}
\eeq  
which depends on isospin since $N\neq Z$ in all the cases of interest.
The exchange part, which takes into account the effect of antisymmetrization 
under the exchange of nucleons between the two nuclei, is modeled by the 
contact interaction, 
\beq
v_{\rm ex}(\bd{r}_{12})=\left ( {\hat J}_{\rm 00}+
\frac{N_1-Z_1}{A_1} \ \frac{N_2-Z_2}{A_2} \ {\hat J}_{01}\right ) \ 
\delta(\bd{r}_{12}).
\label{vex}
\eeq

The density independent part of the effective nucleon-nucleon force that 
we use is the Reid parametrization of Michigan-3-Yukawa (M3Y) interaction 
\cite{bert77}.  The explicit form of the expressions for $v_{00}, v_{01}, 
{\hat J}_{00}$ and ${\hat J}_{01}$ are given, for example,
in Ref. \cite{sand99}. 
Moreover, we neglect the possible energy dependence of these parameters. 
For example, a variation  $\delta E\approx 25$ MeV between the minimum and
maximum bombarding energies considered in the fusion of $^{64}$Ni+$^{64}$Ni 
gives a variation of 
$\delta {\hat J}_{00}\approx 0.27~{\rm MeV} \cdot{\rm fm}^{3}$ MeV in the Reid 
parametrization, a values that is small compared to 
${\hat J}_{00}\approx -276~{\rm MeV} \cdot{\rm fm}^{3}$. 

As we noted earlier, a potential that is based on the M3Y interaction 
predicts correctly the ion-ion potential for peripheral collisions. 
However, reactions that are sensitive to the potential at smaller 
distances are not reproduced \cite{brasat97}.  To cure this deficiency,
the ion-ion potential is supplemented with a short-ranged repulsive 
potential which, according to the discussion in the previous subsection, 
is proportional to the overlapping volume of the reacting nuclei.
This is simulated in Eq.~(\ref{dfold}) by using
the effective contact interaction, 
\beq
v_{\rm rep}(\bd{r}_{12}) = V_{\rm rep} \ \delta(\bd{r}_{12}).
\label{vrep}
\eeq
We follow the procedure proposed in Ref. \cite{ueg93} and use a relatively
sharp density profile, characterized by the diffuseness $a_{\rm rep}$, when 
calculating the repulsive potential from the double-folding integral, 
Eq.~(\ref{dfold}).
The strength $V_{\rm rep}$ of the repulsive interaction is then obtained 
from the condition Eq.~(\ref{DVCAL}), which relates the total nuclear 
potential at the coordinate origin to the nuclear incompressibility.

\subsection{Parametrization of densities}

The densities that appear in Eq.~(\ref{dfold}) are parametrized with 
Fermi-Dirac distribution functions
\beq
\rho_i(\bd{r})=\frac{\rho_0}{1+\exp[(r-R_{0i})/a]}, 
\label{dens}
\eeq
where \ $R_{0i}=r_{0i}A_i^{1/3}$.
The parameters of the proton and neutron density distributions 
we have used are listed in Table \ref{tabla1}.
\begin{table}
\begin{center}
\begin{tabular}{|c|c|c|c|c|}
\hline
Nucleus  & $r_{0p}$ (fm) & $a_p$ (fm) & $r_{0n}$ (fm)& $a_{n}$ (fm) \\
\hline
 $^{58}$Ni& 1.107      & 0.4673  & 1.0836 & 0.5124      \\
\hline
 $^{64}$Ni& 1.0652     & 0.575   & 1.0852 & 0.532      \\
\hline
$^{100}$Mo& 1.1025     & 0.535   & 1.105  & 0.575 \\
\hline
\end{tabular}
\caption{Parameters of the proton and neutron density distributions,
Eq. (\ref{dens}), which have been used in the calculation of the direct
and exchange, double-folding potential.}
\label{tabla1}
\end{center}
\end{table} 
The parameters for the proton density of $^{64}$Ni are taken 
from a compilation of elastic electron scattering  data \cite{vries87}.
For the neutron distribution we choose 
the parameters such that they are in the range of what one would expect for 
a moderately neutron-rich nucleus, and such that the barrier of the 
M3Y+repulsion potential is close to the one predicted by the 
Aky\" uz-Winther potential, since this potential gives a good description 
of the data in the barrier and high-energy region \cite{jiang04b}. 
The parameters for the proton and neutron densities in $^{58}$Ni and $^{100}$Mo 
we have used are identical or close to those predicted by the 
Hartree-Fock-Bogoliubov (HFB) method using the BSk2 Skyrme force \cite{ripl-2}.

\begin{table}
\begin{center}
\begin{tabular}{|c|c|c|c|c|}
\hline
Reaction  & $a_{\rm rep}$ (fm) & $V_{\rm rep}$ (MeV$\cdot$fm$^3$) & $K$ (MeV) & 
$V_{\rm pocket}$ (MeV) \\
\hline
 $^{58}$Ni+$^{58}$Ni & 0.341      & 510.1  & 234   & 89.31       \\
\hline
 $^{64}$Ni+$^{64}$Ni & 0.405      & 511.0  & 228   &  84.98        \\
\hline
 $^{64}$Ni+$^{100}$Mo& 0.375      & 488.7  & 226  & 117.8    \\
\hline
\end{tabular}
\caption{Parameters used in the calculation of the repulsive, double-folding 
potential for three fusion reactions, and the associated incompressibility 
and pocket energy.}
\label{tabla2}
\end{center}
\end{table} 
\begin{figure}[t]
\center{\epsfig{figure=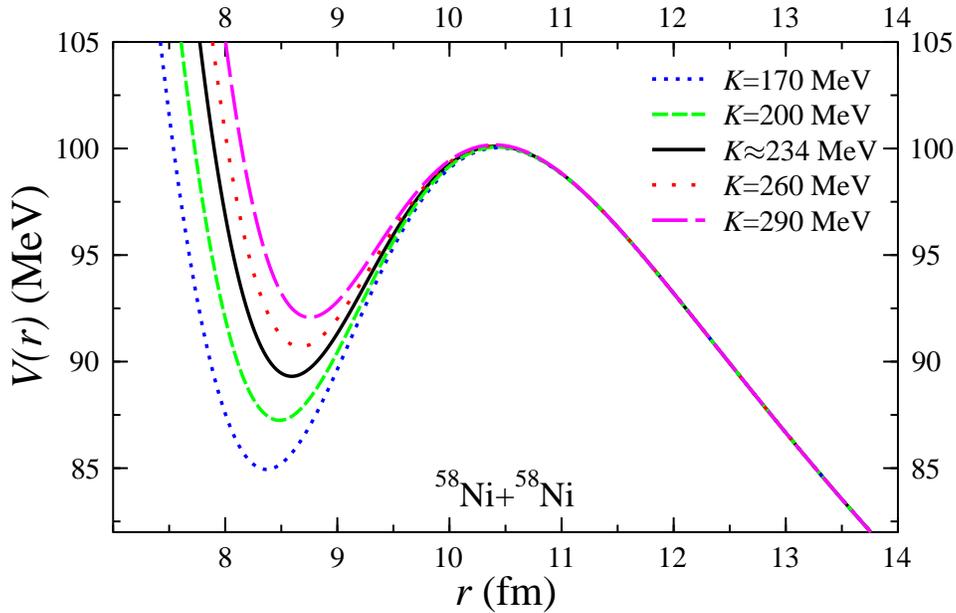,width=0.8\textwidth}}
\caption{(Color online). 
Coulomb plus the nuclear M3Y+repulsion, double-folding potential for 
$^{58}$Ni+$^{58}$Ni. Results are shown for different values of the nuclear 
incompressibility $K$.}
\label{potkdep_ni58_ni58}
\end{figure}

\subsection{Heavy-ion potential study cases}

Applying the formalism described in subsections A and B, we vary the parameters 
$a_{\rm rep}$ and $V_{\rm rep}$ of the repulsive potential so that 
Eq. ~(\ref{DVCAL}) is approximately fulfilled for the nuclear incompressibility 
$K$ predicted by the Thomas-Fermi model \cite{myswy98}.
The parameters we obtain for the three heavy-ion systems we have studied are
listed in Table \ref{tabla2}, together with the incompressibility of the compound 
nuclei obtained in the Thomas-Fermi model, and the energy of the 
resulting pocket that appears in the Coulomb plus nuclear potential.
 
To illustrate the dependence of the total ion-ion potential on the nuclear 
equation of state we plot in Fig.~\ref{potkdep_ni58_ni58} the Coulomb plus 
nuclear potential for the dinuclear system $^{58}$Ni+$^{58}$Ni
for different choices of the nuclear incompressibility $K$.
The solid black curve corresponds to the incompressibility $K\approx$ 234 MeV 
as inferred from the Thomas-Fermi model for $^{116}$Ba. 
The effect of $K$ on the inner part of the barrier and the depth of the
pocket is essential.  A soft EOS provides a deep pocket whereas a hard one 
results in a shallow pocket. 

\begin{figure}
\center{\epsfig{figure=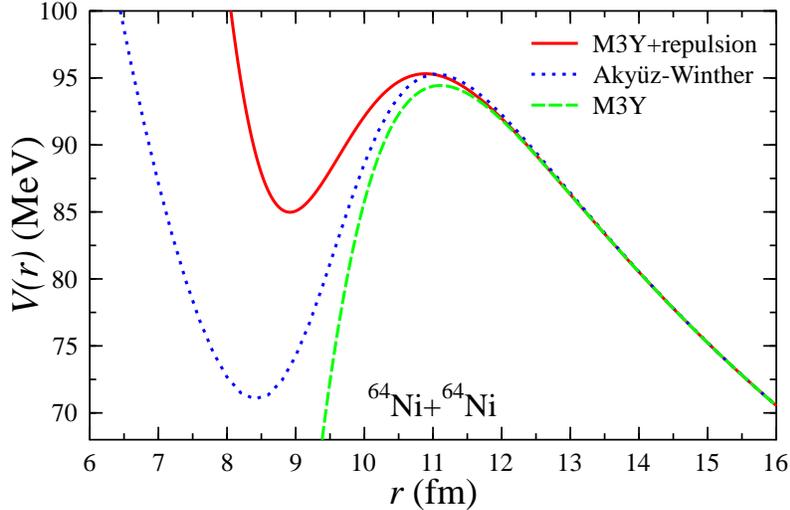,width=0.7\textwidth}}
\caption{(Color online).
Comparison of the double folding potentials with (solid curve)
and without (long-dashed curve) repulsion, and the Aky\"uz-Winther potential 
(short dashed curve) for the system $^{64}$Ni+$^{64}$Ni.}
\label{reid_ni64_ni64}
\end{figure}

The effect of the repulsive core is displayed in Fig.~\ref{reid_ni64_ni64}.
We compare the total potentials obtained from standard M3Y heavy-ion potential,
the M3Y+repulsion potential for $^{64}$Ni+$^{64}$Ni, and the
the Aky\"uz-Winther, for the system $^{64}$Ni+$^{64}$Ni. 
Due to the lower barrier and an abrupt decrease in the inner region, the 
M3Y potential cannot accurately reproduce the data, a fact that already has been 
pointed out in the literature \cite{new04,gont04}. 

\begin{figure}
\centering
\mbox{\subfigure{\epsfig{figure=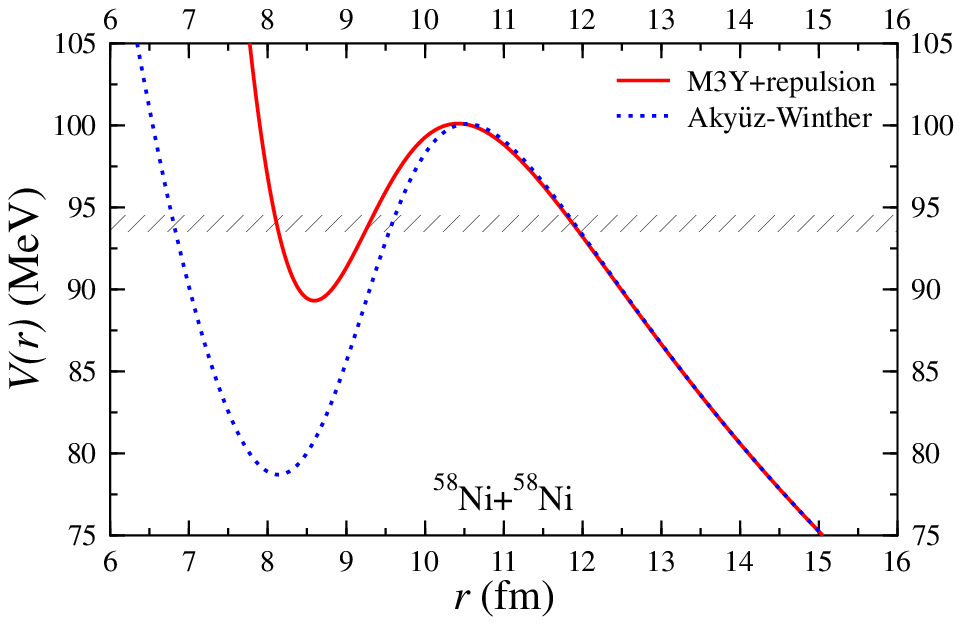,width=0.5\textwidth}}
      \subfigure{\epsfig{figure=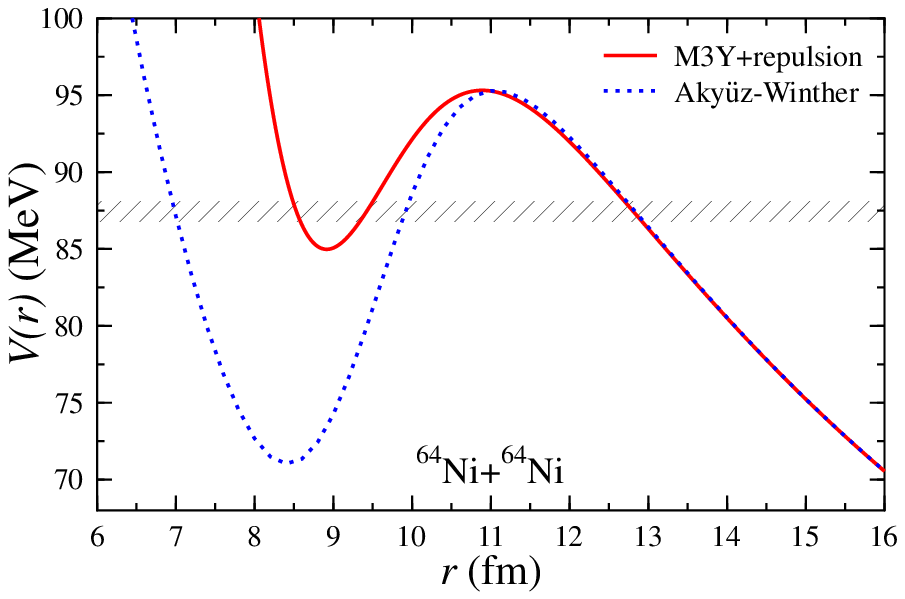,width=0.5\textwidth}}
     }
\mbox{\subfigure{\epsfig{figure=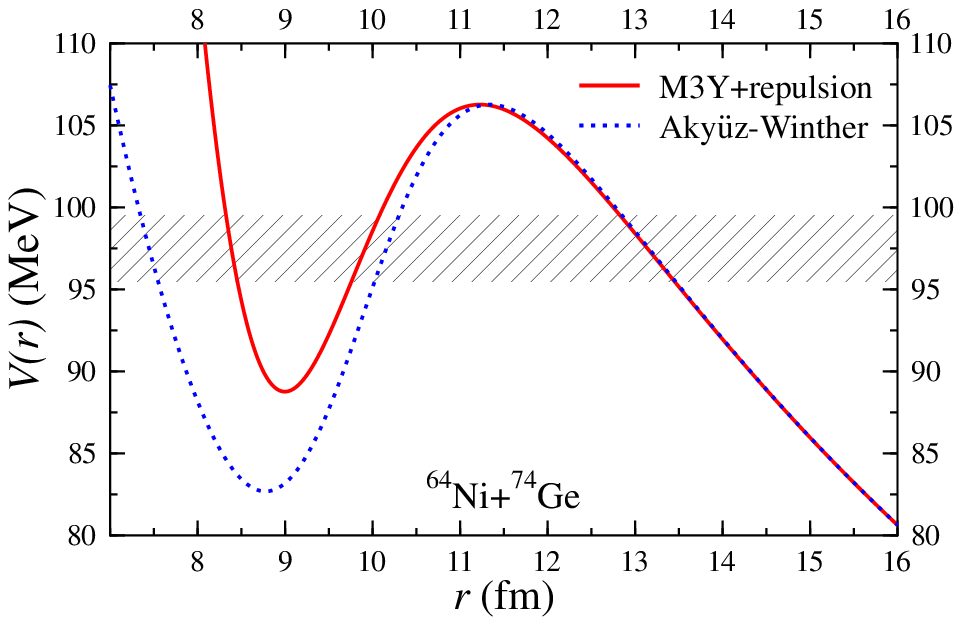,width=0.5\textwidth}}
      \subfigure{\epsfig{figure=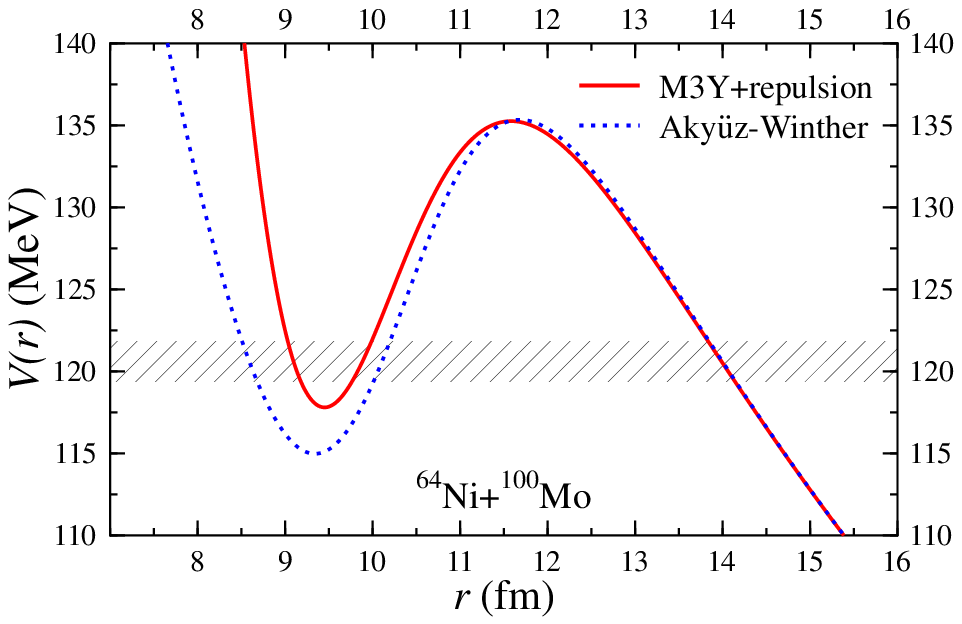,width=0.5\textwidth}}
      }
\caption{(Color online).
Ion-ion potentials for $^{58}$Ni+$^{58}$Ni, 
$^{64}$Ni+$^{64}$Ni, $^{64}$Ni+$^{74}$Ge, $^{64}$Ni+$^{100}$Mo.
The solid curves are the potentials employed in the present work.
The short-dashed curves are based on the Aky\"uz-Winther potential, 
which was used in Refs. \cite{jiang02,jiang04a,jiang04b,jiang05,jiang06}. 
The dashed strips show for each system the experimental boundaries 
of the threshold energy $E_s$ \cite{jiang06}.}
\label{potentialfigs}
\end{figure}

In Fig.~\ref{potentialfigs} we show the profiles of the Coulomb plus nuclear potentials 
for four heavy-ion systems.
The solid curves are based on the M3Y+repulsion potentials we use in this 
work, whereas the short-dashed curves are based on the AW potential.  
It is seen that the pockets predicted for the M3Y+repulsion potentials 
are much shallower than predicted by the AW potential, in particular for 
the lighter systems. 
The dashed region in Fig.~\ref{potentialfigs} shows the energy $E_s$ where 
the experimental $S$-factor develops a maximum, and the width of it
illustrates the uncertainty in the extracted value of $E_s$ \cite{jiang06}.

\section{Coupled-Channels Approach}

We use the same approach as in previous publications (see \cite{esb04} and references 
therein), i.e. coupled-channels calculations performed in the so-called iso-centrifugal 
or rotating-frame approximation, where it is assumed that the orbital angular momentum 
$L$ for the relative motion of the dinuclear system is conserved. 
The rotating frame approximation allows a drastic reduction of the number of channels 
used in the calculations. 

The only ingredient we are going to change in the formalism is the ion-ion potential.  
Like in the case of other nuclear potentials we resort to the "proximity" approximation 
\cite{browin91} which states that the heavy-ion potential is a function of the shortest 
distance between the nuclear surfaces of the reacting nuclei. 
Since for spherical nuclei, the relative orientation is not coming into play, the 
fragments have the natural tendency to be polarized along the collision axis,
which is defined by the radial separation vector $\bd{r}$.
Surface vibrations are then taking place parallel (along) the direction $\bd{r}$. 
In this case the shortest distance between the two surfaces is
\beq
s=r-R_1-R_2-\delta R~,         
\eeq
where
\beq
\delta R=R_1\sum_{\lam\mu}\alpha_{\lam\mu}^{(1)}Y_{\lam\mu}^{*}(\hat r)
        +R_2\sum_{\lam\mu}\alpha_{\lam\mu}^{(2)}Y_{\lam\mu}^{*}(-\hat r), 
\label{surfdist}
\eeq
and $\hat r$ specifies the spatial orientation of the dinuclear system in the
laboratory system and $\alpha_{\lam\mu}^{(i)}$ are the deformation parameters.
In the rotating frame approximation, which we will use, the direction of $\bd{r}$ 
defines the $z$-axis. The only vibrational excitations that can take place are 
therefore the $\mu=0$ components, since $Y_{\lam\mu}(\hat z)\propto\delta_{\mu,0}$. 

The cases under study in this paper refer only to the generation of vibrational 
excitations of surface modes.  In previous papers it was shown that the 
inclusion of linear and quadratic interactions is necessary and often 
sufficient to fit the data at least in the intermediate energy region 
(see \cite{esb04} and references therein).
It is then necessary to consider only the spherical part of the double-folding 
potential, and its first and second order derivatives. 
The nuclear potential for the elastic channel is then given by Eq. (\ref{dfold}),
which is most conveniently calculated by the Fourier transform \cite{misgre02},
\beq
V(r)=\frac{1}{2\pi^2}
\int dq~q^2~ \rho_{1}(q)~\rho_2(q)~v(q)~j_{0}(qr).
\label{potmon}
\eeq
Here $\rho_i(q)$ is the Fourier transform of the spherical ground 
state density of ion $i$, $v(q)$ the Fourier transform of the effective 
$NN$ interactions, and $j_0(qr)$ stands for the spherical Bessel function.
This is actually the potential we have already shown in Figs. 1-3.

The non-spherical  part of the nuclear potential that results from the difference 
between the total interaction and the potential in the elastic channel is expanded 
up to second-order in the surface distortion (\ref{surfdist}),
\beq
\delta V_N(r)= -\frac{\partial V}{\partial r} \delta R 
+\frac{1}{2}\frac{\partial^2 V}{\partial r^2} \
[ (\delta R)^2 - \langle 0| (\delta R)^2 |0\rangle ].
\eeq
It is seen that the ground state expectation of this interaction,
$\langle 0| \delta V_N | 0\rangle$, is zero, but the second order term will 
give a non-zero contribution to the diagonal matrix element in an excited state.
One can show that this prescription is exact for a harmonic oscillator up to 
second order in the deformation amplitudes.
We include a similar expansion of the Coulomb field, $\delta V_C$, but only 
to first order in the deformation amplitudes \cite{esb04}.
These expressions are inserted into the C.~C. formalism in the rotating frame 
approximation which singles out only axially symmetric 
distortions ($\alpha_{\lambda\mu=0}$), i.e.,
$$
\left ( \frac{\hbar^2}{2M_0} \left [-\frac{d^2}{dr^2} + \frac{L(L+1)}{r^2}\right ]
+ \frac{Z_1Z_2e^2}{r}
+ V(r)+\sum_{n_1,n_2} \varepsilon_{n_1,n_2}-E\right )
u_{n_1n_2}(r)
$$

\beq 
= -\sum_{m_1m_2}\langle n_1 n_2\mid \delta V_C + 
\delta V_N \mid m_1 m_2\rangle u_{m_1m_2}(r),
\label{ccequ}
\eeq
where $E$ is the relative energy in the center of mass frame, 
$L$ is the conserved orbital angular momentum, and $M_0$ is the
reduced mass of the dinuclear system. 
The C.~C. equations (\ref{ccequ}) are written for two coupled vibrators
of eigenenergy $\varepsilon_{n_1,n_2}$ and consequently the radial wave 
function $u(r)$ is labeled by the quantum numbers $n_1$ and $n_2$. 
Expressions for the matrix elements of $\delta V$ in the double-oscillator 
basis are given in \cite{esb-land87}. 

The above C.~C. equations are solved with the usual boundary conditions 
at large distances and appropriate in-going-wave boundary conditions 
imposed inside the barrier, more precisely, at the radial separation 
where the potential pocket attains its minimum. The fusion cross section  
results then from the total in-going flux.
The calculations include one-phonon excitations of the lowest
$2^+$ and $3^-$ states in target and projectile, and all two-phonon
and mutual excitations of these states up to a 7.2 MeV excitation energy.
This energy cutoff was chosen so that all of the two-phonon states were 
included in the calculations for $^{64}$Ni+$^{64}$Ni and $^{64}$Ni+$^{100}$Mo, 
whereas the two-phonon octupole states were excluded in the calculations
for $^{58}$Ni+$^{58}$Ni.
The necessary structure input for $^{64}$Ni and $^{100}$Mo is given in Ref. 
\cite{esb05}; the input for $^{58}$Ni is from Ref. \cite{esb-land87}.

\section{Cross-sections, $S$-factors and logarithmic derivatives}

In what follows we analyze the fusion data for three systems 
for which there are strong indications of a hindrance, -or steep 
falloff of the cross sections, at energies well below the Coulomb barrier.
For completeness we also show in section VI.4 the high-energy behavior 
of the cross sections.  All calculations are based on the coupled-channels 
formalism outlined in the previous section, and further interpretations 
of the results are presented in Sect. VII. 

\subsubsection{$^{58}$Ni+$^{58}$Ni}

\begin{figure}[t]
\center{\epsfig{figure=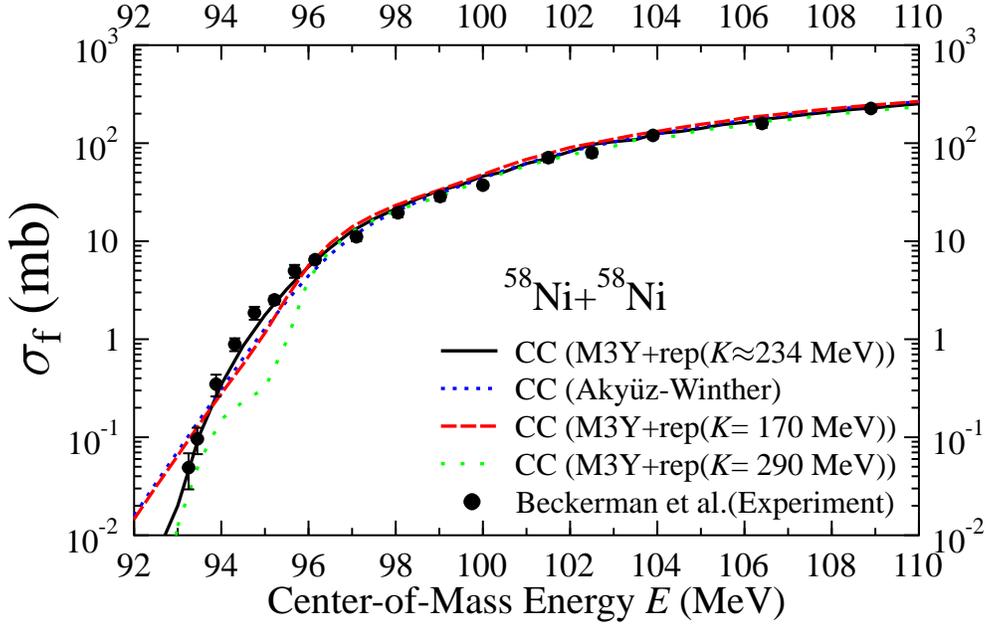,width=0.8\textwidth}}
\caption{(Color online).
Experimental fusion excitation function for the system 
$^{58}$Ni+$^{58}$Ni \cite{beck82} is compared to various C.~C. calculations 
that are based on the M3Y+repulsion potential (assuming different values
of the nuclear incompressibility $K$), and on the AW potential.}
\label{sig_ni58_ni58}
\end{figure}

The first case we consider is the fusion reaction of $^{58}$Ni+$^{58}$Ni.
For this system  data are available from an older experiment 
and the smallest cross sections are in the range of mb \cite{beck82}. 
The data are compared in Fig.~\ref{sig_ni58_ni58} to several C.~C. 
calculations that are based on the 
same structure input as in Ref. \cite{esb-land87} and on the
M3Y+repulsion potential, but they
differ in the assumed value of the nuclear incompressibility.
We see that for a hard nuclear EOS ($K$=290 MeV) the calculated 
low-energy cross sections deviate strongly from the measured values.
Obviously in this case the origin of the mismatch is primarily due to the 
existence of a very shallow pocket which reduces the classically allowed 
region and hinders the fusion drastically. 
For a soft nuclear EOS ($K$=170 MeV) the potential pocket is lowered 
and the calculated cross sections approach the results calculated with the 
Aky\"uz-Winther potential.  The best  fit to the data is obtained for the
nuclear incompressibility $K$=234 MeV, which is the value one obtains
in the extended Thomas-Fermi model. 

\begin{figure}[t]
\center{\epsfig{figure=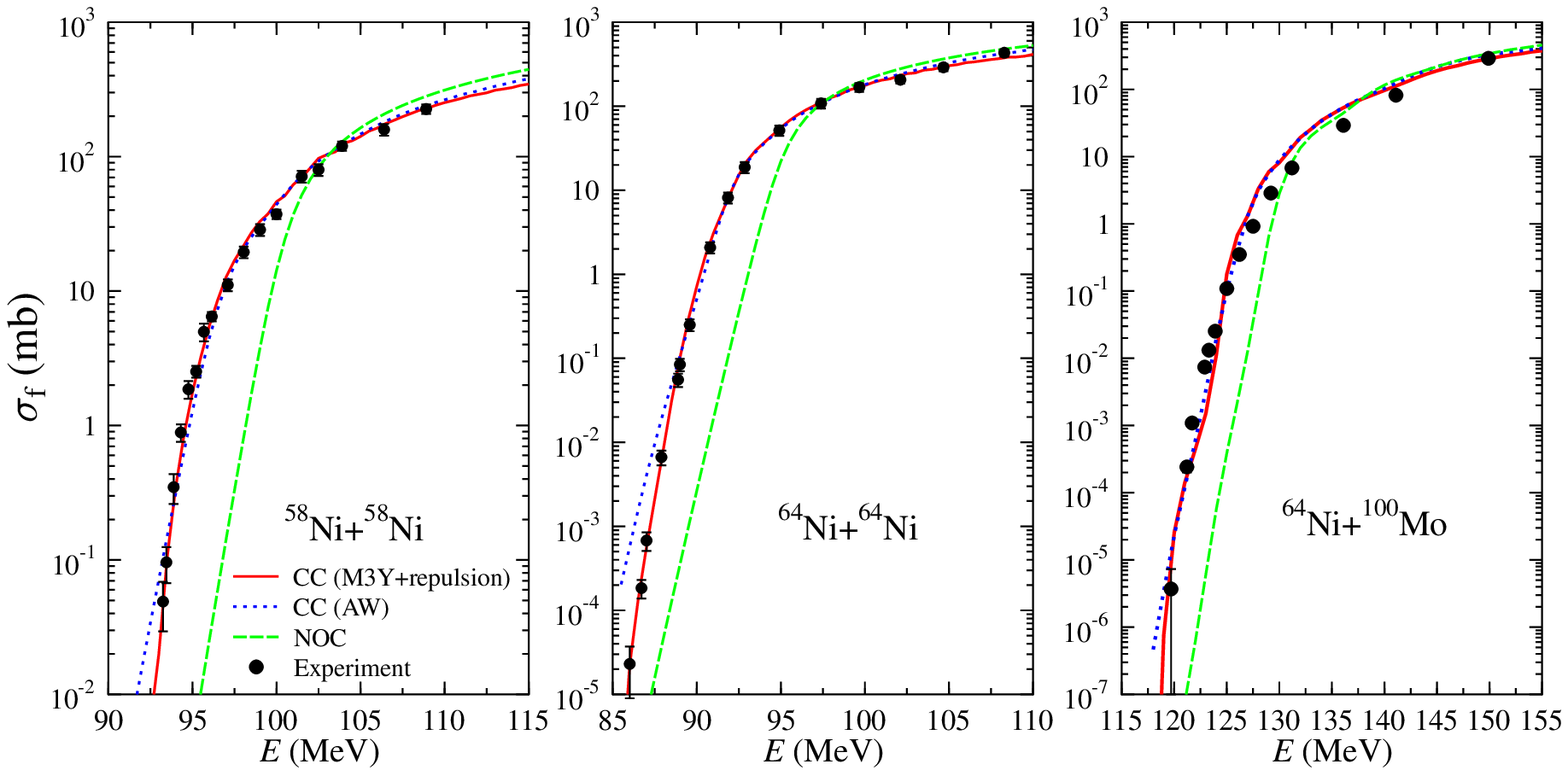,width=1.0\textwidth}}
\caption{(Color online). 
Experimental fusion excitation functions for the systems 
$^{58}$Ni+$^{58}$Ni \cite{beck82}, 
$^{64}$Ni+$^{64}$Ni \cite{jiang04b}, 
and $^{64}$Ni+$^{100}$Mo \cite{jiang05}
are compared to various C.~C. calculations described in the text,
and to the no-coupling (NOC) limit for the AW potential.}
\label{sigmas}
\end{figure}

Some of the results shown Fig.~\ref{sig_ni58_ni58} are repeated in
Fig.~\ref{sigmas}, where they are compared to similar results
for the fusion of $^{64}$Ni+$^{64}$Ni and $^{64}$Ni+$^{100}$Mo,
which will be discussed in more detail below. The C.~C. calculations 
for the M3Y+repulsion potential (solid curves) were all performed with
the nuclear incompressibility derived from the extended Thomas-Fermi 
model. The long-dashed curves are the no-coupling limits (NOC) 
obtained with the AW potential.

\begin{figure}[t]
\center{\epsfig{figure=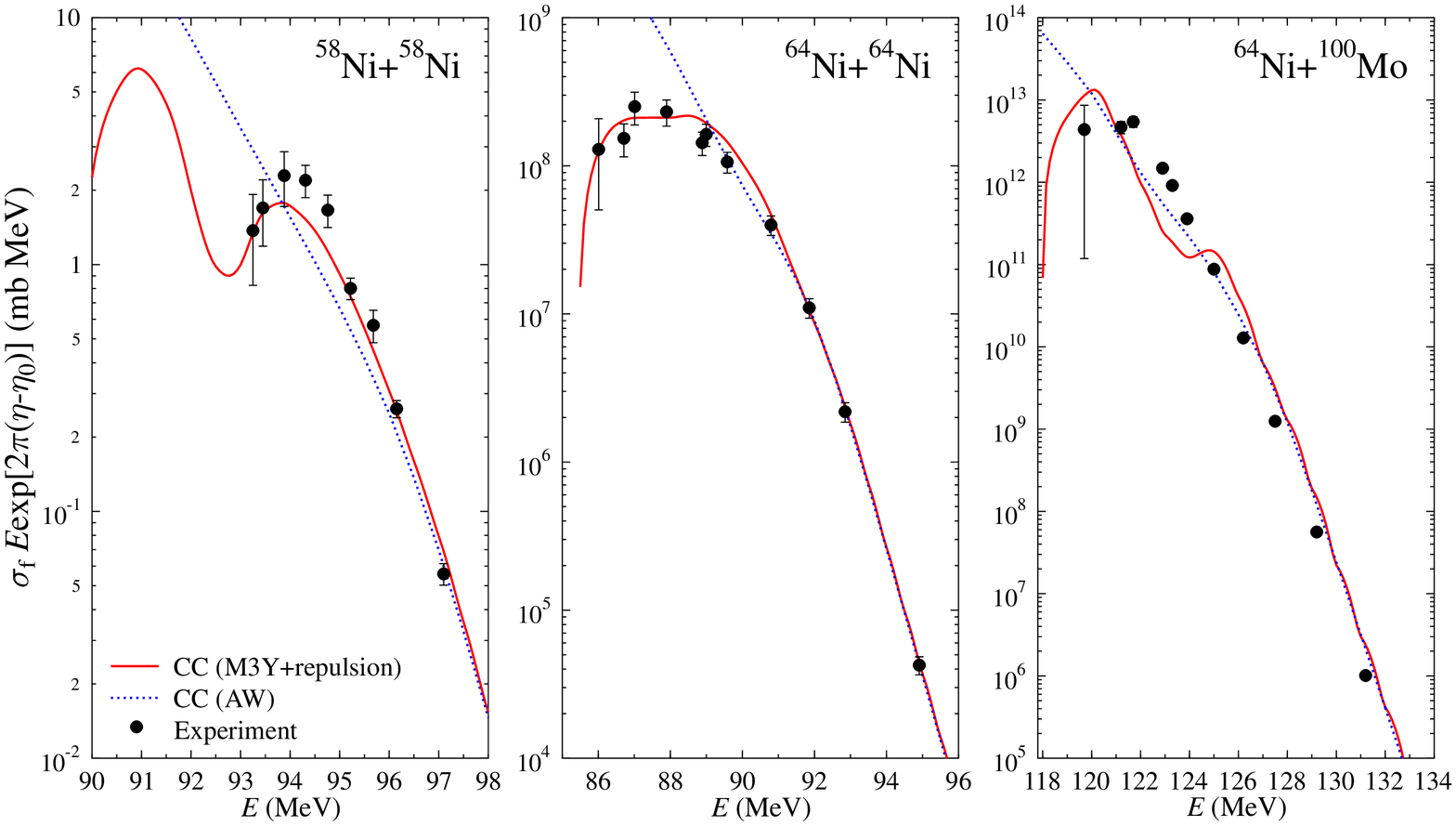,width=1.0\textwidth}}
\caption{(Color online).
The experimental $S$-factors for the systems 
$^{58}$Ni+$^{58}$Ni \cite{beck82} (left panel), 
$^{64}$Ni+$^{64}$Ni \cite{jiang04b} (middle panel) and 
$^{64}$Ni+$^{100}$Mo \cite{jiang05} (right panel) are indicated by solid circles.
They are compared to the coupled-channels calculations performed with the 
M3Y+repulsion (solid curve) and Aky\"uz-Winther (dashed curve) potentials.}
\label{sfa_factors}
\end{figure}

In Fig.~\ref{sfa_factors} (left panel) we compare the experimental $S$-factors
for $^{58}$Ni+$^{58}$Ni to the those calculated with the M3Y+repulsion and the
AW potentials. In the last case the mismatch with the data is already evident 
starting at energies 4-5 MeV below the barrier. The former case, on the
other hand, provides a good description of the available  data and,
most importantly, reproduces the trend of the experimental $S$-factor 
to produce a maximum. The calculation also predicts a second maximum at even 
lower energies where no data exist. It remains to be seen whether 
future measurements will confirm the two predicted maxima.

\begin{figure}
\center{\epsfig{figure=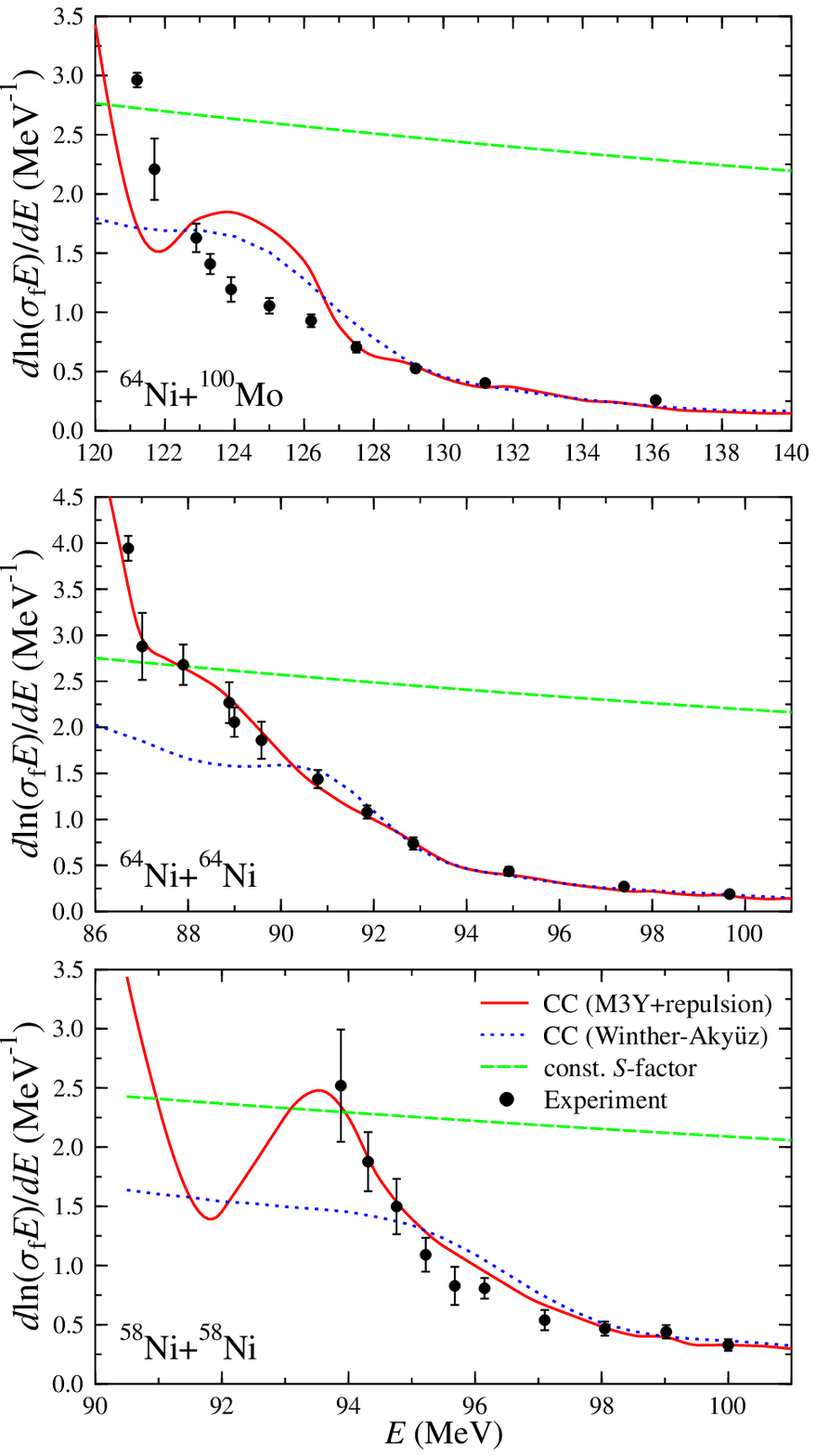,width=0.6\textwidth}}
\caption{(Color online).
Logarithmic derivatives of the energy-weighted cross sections 
for $^{58}$Ni+$^{58}$Ni (lower panel), $^{64}$Ni+$^{64}$Ni (middle panel), 
and $^{64}$Ni+$^{100}$Mo (upper panel). Experimental results
derived from the cross sections shown in Fig. \ref{sigmas} are compared 
to coupled-channels calculations performed with the M3Y+repulsion (solid curve) 
and the Aky\"uz-Winther (dashed curve) potential.
The top dashed curve in each panel is the prediction for a constant $S$-factor.}
\label{log_derivat}
\end{figure}

In Fig.~\ref{log_derivat} (lower panel) we plot the logarithmic derivative 
of the energy-weighted cross section for the case $^{58}$Ni+$^{58}$Ni.
We obtain  a nice description of the available experimental data when we use
the M3Y+repulsion potential (solid curve). As expected from the behavior 
of the $S$-factor, $L(E)$ produces a maximum (just above the constant $S$-factor 
limit) followed by a local minimum, before it finally diverges at lower energies.
To ascertain this behavior, and also the predicted excitation function
and $S$-factor for $^{58}$Ni+$^{58}$Ni, high-precision fusion data are 
necessary at very low bombarding energies,

\subsubsection{$^{64}$Ni+$^{64}$Ni}

Next we consider the system $^{64}$Ni+$^{64}$Ni. This case was already
discussed in our previous publication \cite{misesb06a} but for the sake of 
completeness we find it necessary to recall it and add more features like,
for example, the logarithmic derivative.

The system $^{64}$Ni+$^{64}$Ni was advocated to be a very good choice for 
precise measurements 
since there is no contamination from reactions with heavier isotopes in 
the target, or from lower-$Z$ isobaric components in the beam, which can 
dominate the yield at extreme sub-barrier energies \cite{jiang04b}. 
The ATLAS data for this case are in a good agreement with previous results 
\cite{acker96} at energies above the Coulomb barrier but not around and 
below the barrier. However, the agreement is good around and below the 
barrier with an older experiment \cite{beck82}.
It should be also mentioned that the ATLAS data provide 
cross sections down to tenths of nanobarns \cite{jiang04a} compared to
the 0.3 mb that was reached in the older experiment \cite{beck82}.

To obtain the incompressibility of $K\approx$ 228 MeV for total overlap of 
the reacting nuclei, as predicted by the Thomas-Fermi model, we use a 
strength  of $V_{\rm rep}$ = 511 MeV and a diffuseness of the repulsive 
part of the density distribution of $a_{\rm rep}\approx$ 0.4 fm for both 
protons and neutrons. 

In Fig.~\ref{sigmas} (middle panel) we compare the experimental excitation 
function for the fusion reaction $^{64}$Ni + $^{64}$Ni $\rightarrow ^{128}$Ba 
with the C.~C. results obtained using the Aky\" uz-Winther potential like in 
Ref. \cite{jiang02} (dotted line) and with the M3Y+repulsion potential 
(solid line). The same recipe for the C.~C. calculations was used in both cases.
The diffuseness of the Aky\" uz-Winther potential we use is $a$ = 0.676 fm. 
One should also recall that calculations using a modified Aky\"uz-Winther 
potential, where the diffuseness inside the barrier was set to $a_i$ = 5 fm, 
were not able to the date to explain the steep fall-off of the measured
cross section \cite{jiang04a}. This is not surprising because the modified 
AW potential has a very deep pocket (see Fig. 4 of Ref. \cite{esb04} for an 
interior diffuseness of $a_i$ = 10 fm.)

We conclude, from the inspection of Fig.~\ref{sigmas} that the agreement 
with data, when using the M3Y+repulsion potential, is sensitively better 
than the one provided by the Aky\" uz-Winther starting at 90 MeV, and not 
only for the 4 lowest experimental data points. 
The excitation function obtained with the M3Y+repulsion potential has the 
right slope, not only because the potential attains a higher-lying pocket 
but also because the curvature of the barrier is different, with a thicker 
barrier in the overlapping region.   
Thus the best $\chi^2$ per point is only $\chi^2/N$ =  0.86. This value
is obtained by applying the energy shift $\Delta E$ = 0.16 MeV to the 
calculated excitation function. 
The best fit obtained with AW potential, on the other hand, gives a 
$\chi^2/N$ = 10 and requires an energy shift of $\Delta E$ = 0.9 MeV.

The $S$-factor representation of the $^{64}$Ni+$^{64}$Ni fusion cross 
sections is displayed in the middle panel of Fig.~\ref{sfa_factors}.
The clear maximum in the measured $S$-factor is reproduced only by the 
M3Y+repulsion potential. At this point we recall the experience gained in 
the past on molecular resonances. As shown in Ref. \cite{pat71}, the 
$S$-factor exhibits a sequence of quasi-molecular resonances, sandwiched 
between a limiting interior threshold and the Coulomb barrier.
In the present case we obtain a maximum that is too broad to be assigned 
to a resonance, the curvature in the $S$-factor being explained by the 
shallow pocket in the potential. The maximum of the theoretical curve 
corresponds approximately to the maximum of the experimental data.

The logarithmic derivative of the energy-weighted cross section is shown 
in the middle panel of Fig.~\ref{log_derivat}, and a nice agreement with
the data is obtained in the C.~C. calculations that are based on the 
M3Y+repulsion potential.

\subsubsection{$^{64}$Ni+$^{100}$Mo}

For the repulsive part of the potential we choose 
$a_{\rm rep}=0.375$ fm and $V_{\rm rep}$ = 488.7 MeV, 
values that are matching the incompressibility 
$K$ = 226 MeV of the compound nucleus $^{164}$Yb.
To improve the fit to the data we included up to three phonon excitations 
of the quadrupole mode in $^{100}$Mo using the structure parameters
given in Ref. \cite{esb05}. 
However, the agreement with the data seen in the right panel of 
Fig.~\ref{sigmas} is clearly not as good as in the other two cases shown 
in the same figure. The reason is that the C.~C. effects are very strong
for this heavy-ion system and the calculations have not fully converged 
with respect to multiphonon excitations, as discussed in Ref.~\cite{esb05}.
Another problem is that the nuclear structure properties of multiphonon
states are often poorly known, so we will not try to improve the fit to
the data here.

For the $S$-factor we reproduce roughly, as can be concluded from an 
inspection of the right panel in Fig.~\ref{sfa_factors}, the trend to 
develop a well pronounced maximum at the lowest measured energies and 
provide a reasonable estimation for this maximum.
Also the logarithmic derivative manifests the tendency to develop a divergent
behavior at low energies (Fig.~\ref{log_derivat}, upper panel).
The appearance of a local maximum at 124 MeV, on the other hand, is most likely 
caused by a poor convergence with respect to multiphonon excitations.

\subsubsection{High-energy behavior of fusion cross sections}

\begin{figure}[t]
\center{\epsfig{figure=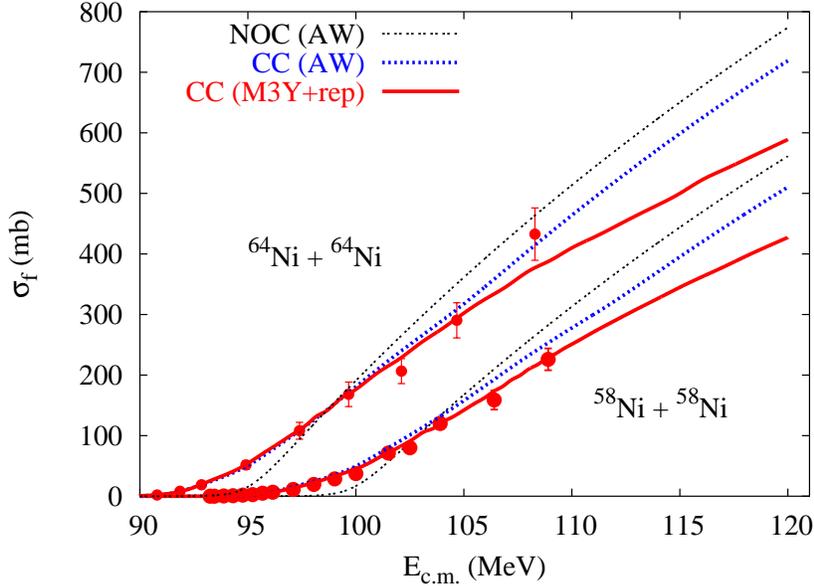,width=0.7\textwidth}}
\caption{(Color online).
Linear plot of the fusion cross sections shown in Fig. \ref{sigmas}
for the two systems $^{64}$Ni+$^{64}$Ni \cite{jiang04b} and 
$^{58}$Ni+$^{58}$Ni \cite{beck82}.}
\label{siglf}
\end{figure}

The cross sections we obtain in coupled-channels calculations 
are suppressed when compared to the no-coupling limit.
This can be seen in Fig. \ref{sigmas} but the suppression looks small on 
a logarithmic plot. We therefore show a linear plot in Fig.  \ref{siglf}
of the same cross sections for the two nickel isotopes.
In both cases we see that the coupled-channels calculations that are based 
on the AW potential are suppressed at high energies when compared to the 
no-coupling limit. This is a well-known C. C. effect but the 
suppression is not always large enough to explain the measurements, 
in particular for heavy systems \cite{new04}.

We also see in Fig. 8 that the C. C. results we obtain 
with the M3Y+repulsion potential are suppressed even further, and 
the suppression increases with increasing energy.
This is naively what one would expect with shallower pockets at 
higher angular momenta.
It is unfortunate that the data considered here 
do not reach very large cross sections, so it is difficult to assess 
the significance of the suppression we obtain.
However, the trend is very encouraging because it helps explaining
why the high-energy fusion data considered in Ref. \cite{new04} are 
suppressed so much compared to C. C. calculations
and why it was necessary to use a large diffuseness of the ion-ion 
potential when fitting the data in the no-coupling limit.

\section{Interpretation of the results}

In the previous section we obtained an excellent fit to the
fusion data for the two nickel isotopes in the C. C. calculations 
that were based on the M3Y+repulsion potential and in-going-wave
boundary conditions. Here we discuss the significance of these
results and the implications for other observables.
We have in this connections also tested other types of repulsive cores, 
like the  gaussian one of the type used in \cite{mich73} but we arrived 
at the conclusion that they are inadequate in reproducing the fusion data 
we have investigated here. Thus there appears to be some limitations or 
constraints in constructing the shape of the ion-ion potential inside 
the barrier.

\subsection{Models of fusion}

The description of heavy-ion fusion is often based on the assumption 
that fusion occurs as soon as the Coulomb barrier has been penetrated. 
More precisely, it is expected that the fusion cross section obtained 
from the absorption in a short-range imaginary potential is essentially 
identical to the cross section obtained with in-going-wave boundary 
conditions (IWBC) \cite{land84}.  Moreover, the calculated fusion cross 
section is expected to be insensitive to small variations in the radius 
where the absorption or IWBC are imposed.

In coupled-channels calculations, 
heavy-ion fusion is usually simulated  by IWBC that are imposed at the 
minimum of the potential pocket, as for example in the computer code
CCFULL \cite{ccfull} and also in the present work. The assumptions 
mentioned above have therefore rarely been challenged, and in cases 
where they have been tested, at energies close to the Coulomb barrier, 
they have usually passed the test.
One exception is discussed in Ref. \cite{land87}, where it was noticed 
that the calculated fusion cross section in one particular case showed 
an unacceptable variation with respect to the radius where the IWBC 
were imposed.  This occurred in a case with particularly strong couplings. 
Another example is Ref. \cite{esb05}, where it was pointed out that 
certain unwanted oscillations in the cross sections, obtained at extreme 
subbarrier energies using the IWBC, could be suppressed by including a 
weak, short-range imaginary potential in the calculations. 

In this work we try to demonstrate that the fusion at extreme subbarrier 
energies is sensitive not only to the thickness and the height of the 
Coulomb barrier but also to the minimum of potential pocket, where the 
IWBC are imposed. It is noted that the calculated fusion cross section, 
which is based on the IWBC, will vanish sharply as the energy approaches 
the minimum of the pocket. This sharp behavior is apparently what is 
needed to fit the data at extreme subbarrier energies, at least in the 
case of $^{64}$Ni+$^{64}$Ni.  

\begin{figure}[t]
\center{\epsfig{figure=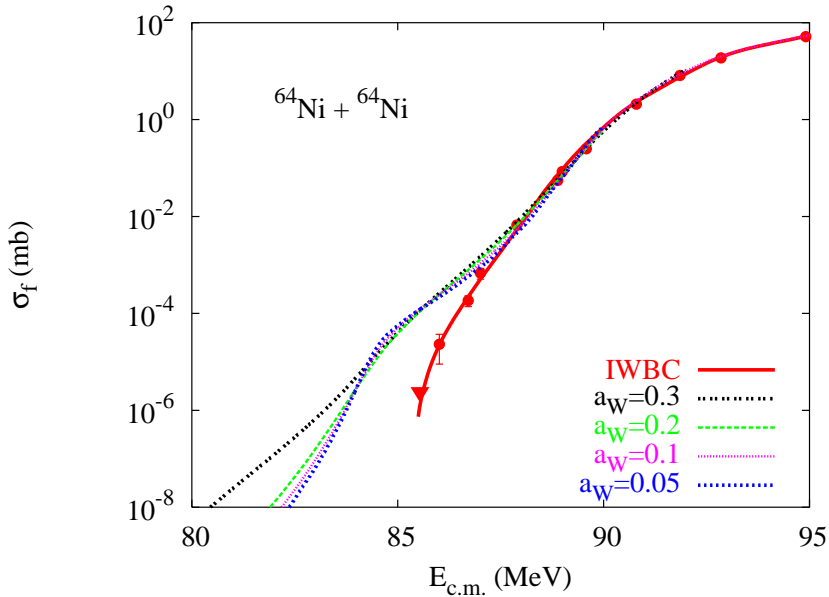,width=0.7\textwidth}}
\caption{(Color online).
Dependence of the calculated fusion cross section (dashed curves)
on the diffusenes $a_W$ of the imaginary potential. The results
are compared to the data for $^{64}$Ni+$^{64}$Ni \cite{jiang04b}
and the calculation, which is based on the IWBC without any absorption 
(the solid curve). The triangular data point at the lowest energy 
is an upper limit.}
\label{figabs}
\end{figure}

We have also calculated the fusion of $^{64}$Ni+$^{64}$Ni using, in
addition to the M3Y+repulsion potential, a short-range imaginary potential 
of the form proposed in Eq. (3) of Ref. \cite{esb05}.
The results for different values of the diffuseness $a_W$  are shown in
Fig. \ref{figabs} by the dashed curves. 
They all make an excellent fit to the data above 10 $\mu$b but they 
deviate significantly from the data at smaller cross sections.
In contrast, the calculation discussed ealier (solid curve), which is 
based on the IWBC without any imaginary potential, is in perfect 
agreement with the data. 

There is a simple reason why the dashed curves are enhanced compared
to the solid curve (IWBC) in Fig. \ref{figabs} at the lowest energies. 
First of all, the fusion obtained from the IWBC without an imaginary 
potential can only occur in the elastic channel at low energy, say 
below 87-88 MeV in the case considered here, and the cross section 
goes sharply to zero at the minimum of the pocket which is 85.4 MeV.
The inelastic channels are energetically closed to fusion at these
low energies and the associated wave functions are (exponentially) 
decaying at the boundary. 
When an imaginary potential is turned on, the fusion/absorption becomes 
possible through all channels. This explains qualitatively why the 
dashed curves are above the solid curve at low energy and why
the conventional assumption about fusion described in the beginning
of this subsection is not correct at extreme subbarrier energies.

\subsection{Average spin for fusion}

\begin{figure}[t]
\center{\epsfig{figure=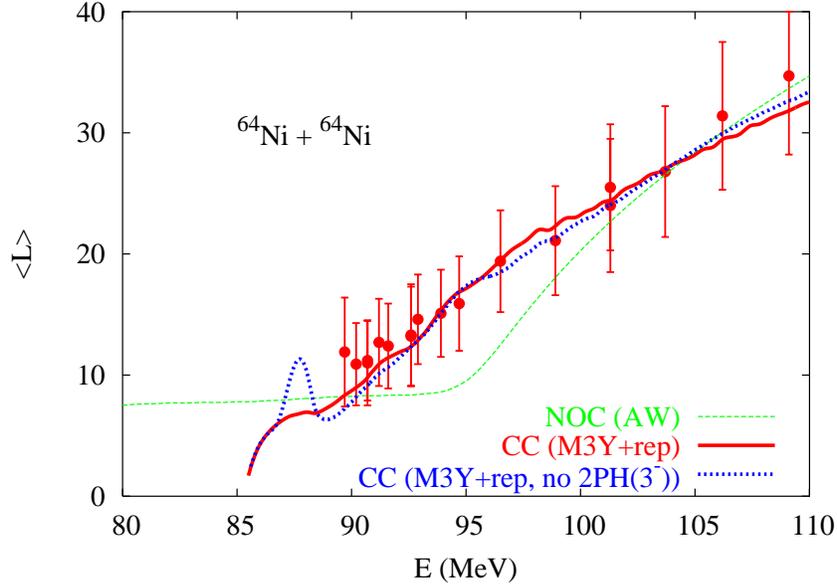,width=0.7\textwidth}}
\caption{(Color online).
Average angular momentum for the fusion of $^{64}$Ni+$^{64}$Ni
obtained in C. C. calculations based on the M3Y+repulsion potential
with (solid curve) and without (thick dotted curve) the effect 
of couplings to the two-phonon octupole states. 
The thin dashed curve was obtained in the no-coupling limit (NOC) 
using the AW potential. The data are from \cite{acker96}.}
\label{avl}
\end{figure}

Another signature of a shallow pocket in the total ion-ion potential is
a narrowing of the spin distribution for fusion as the center-of-mass 
energy decreases and approaches the pocket energy.
In the past it has always been believed that the average angular momentum
for fusion would approach a constant at low energy.
This is the behavior that has been predicted by model calculations, 
including the C. C. \cite{esb04}, but it has not really been tested 
experimentally. An example is shown in Fig. \ref{avl}, where the 
measurements of the $\gamma$-ray multiplicity from the compound nucleus 
formed in the fusion of $^{64}$Ni+$^{64}$Ni
have been converted into an average angular momentum for fusion \cite{acker96}.
The thin dashed curve shows the prediction based on the AW potential
in the no-coupling limit (NOC (AW)). It approaches a constant value
at low energy but the data are always above that limit.
The solid curve in Fig. \ref{avl} shows the results we obtain in 
the C. C. calculations we discussed earlier, which were based on the
M3Y+repulsion potential. It is seen that these calculations predict 
a narrowing of the spin distribution as the energy approaches the 
pocket energy.

After completing our studies we realized that the low-energy behavior 
of several observables can have a strong sensitivity to the couplings 
to multi-phonon states. 
Although the fusion only occurs in the elastic channel at energies 
close to the minimum of the pocket, the polarization of inelastic 
channels can still have a large effect.  We found, in particular, that 
couplings to the two-phonon octupole states are very important. 
This is illustrated in Fig. \ref{avl} by the dotted curve which was 
obtained without any couplings to the two-phonon octupole states. 
It is seen that the calculation in this case develops a rather sharp
peak at 87.7 MeV. The peak disappears when the coupling to the 
two-phonon octupole states is included, as illustrated by the solid curve.
Unfortunately, the data cannot tell us which of these two calculations 
is the most realistic.

\subsection{Structures in calculated cross sections}

\begin{figure}[t]
\center{\epsfig{figure=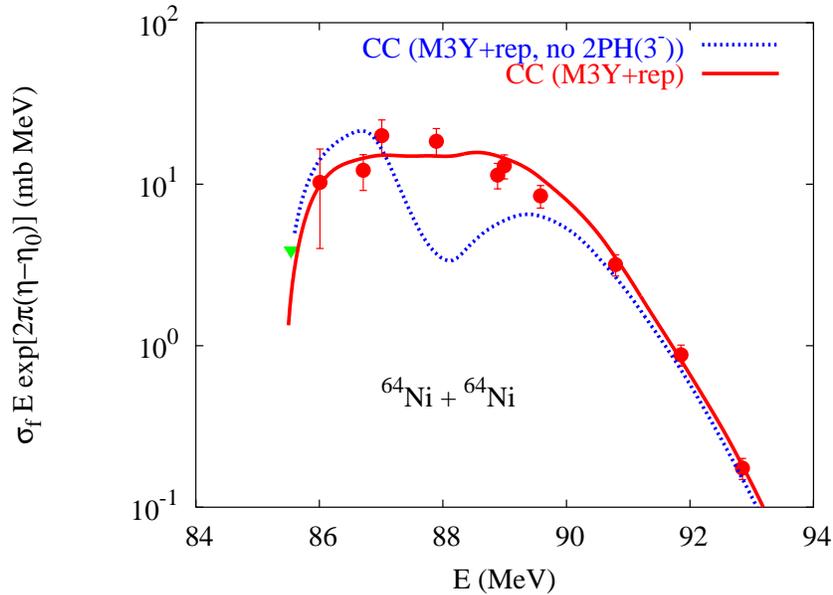,width=0.7\textwidth}}
\caption{(Color online). 
$S$ factors for the fusion of $^{64}$Ni+$^{64}$Ni
obtained in the C. C. calculations based on the M3Y+repulsion 
potential with (solid curve) and without (dotted curve)  
the effect of couplings to the two-phonon octupole states.
The data are from Ref. \cite{jiang04b}; the triangle is an upper limit.}
\label{sfactores}
\end{figure}

The local maximum that appears in the thick dotted curve in Fig. \ref{avl} 
at low energy is not a resonance. It is the result of a vanishing wave 
function in the elastic channel at the radial separation where the IWBC 
are imposed. 
When this condition is fulfilled at low energy, where fusion is 
restricted to the elastic channel, the fusion probability will vanish. 
When it occurs for a range of low angular momenta, it will result in
a large average angular momentum for fusion and that explains the
appearance of the peak in Fig. \ref{avl}.
The suppression of the fusion probability at low energy 
(which occurs when the two-phonon octupole states are not included
in the calculation) can also produce a local minimum in the $S$-factor.
This is illustrated by the dotted curve in Fig. \ref{sfactores}.
The solid curve, which includes the effect of couplings to the 
two-phonon octupole states, shows a single, broad maximum.
In this particular case there is a clear preference for the solid 
curve which makes an excellent fit to the data.

In the calculations we presented for $^{58}$Ni+$^{58}$Ni we did
not include the two-phonon octupole states in the C. C. calculations
because the excitation energy is very large, almost 9 MeV, so we did
not expect these states would be important.
In view of the above discussion it is now understandable why the 
calculated $S$-factor for $^{58}$Ni+$^{58}$Ni shown in left panel 
of Fig. \ref{sfa_factors} develops a double-peaked structure 
at low energy.  We have therefore repeated the calculations and 
included the two-phonon octupole states.  The resulting $S$-factor 
exhibits a single, broad peak, just as we saw in the fusion of 
$^{64}$Ni+$^{64}$Ni.

It is not clear why the couplings to the two-phonon octupole states 
play such an important role as discussed above.
However, the analysis of the $^{64}$Ni+$^{64}$Ni data shows a perfect 
agreement with a single, broad $S$ factor peak, whereas the analysis 
of the existing fusion data for $^{58}$Ni+$^{58}$Ni fusion has a 
strong preference for the double-peaked $S$-factor curve.
In view of these findings it is of interest to continue the fusion 
measurements for $^{58}$Ni+$^{58}$Ni to even lower energies because 
the existing data \cite{beck82} shown in the left panel of Fig.
\ref{sfa_factors} do not verify explicitly the double-peaked structure.

\subsection{Correlation between the pocket energy and $E_s$}

\begin{figure}[t]
\center{\epsfig{figure=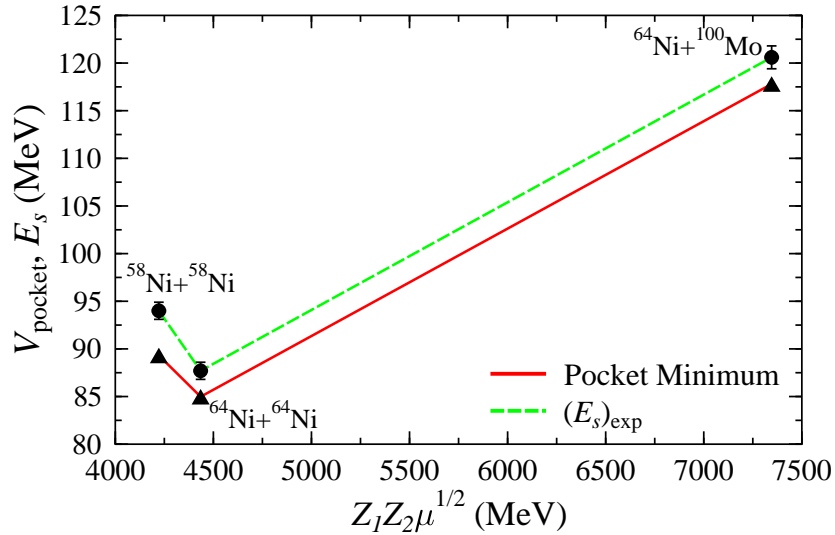,width=0.7\textwidth}}
\caption{(Color online).
Plot of $V_{\rm pocket}$ (triangles) and $E_s$ (solid points) 
versus $Z_1Z_2\sqrt{\mu}$ for the three reactions : $^{58}$Ni+$^{58}$Ni, 
$^{64}$Ni+$^{64}$Ni, $^{64}$Ni+$^{100}$Mo.}
\label{corel_deepfus}
\end{figure}

As we mentioned in the Introduction, Ref.~\cite{jiang04a} reported a correlation 
between the energy $E_s$ where the $S$-factor has a maximum and the parameter 
$Z_1Z_2\sqrt{\mu}$, with $\mu=A_1A_2/(A_1+A_2)$.
In view of the idea of a shallow pocket advocated in 
the present paper we considered it useful to study the dependence 
of the minimum of the potential $V_{\rm pocket}$, along with $E_s$, on the 
parameter $Z_1Z_2\sqrt{\mu}$.
This is illustrated in Fig.~\ref{corel_deepfus} for the three cases we have
investigated in this work. We note the pocket energy $V_{\rm pocket}$ follows 
closely the energy $E_s$, where the $S$ has a maximum, the two results being
separated by about 3 to 5 MeV. 

\section{Conclusions}

We conclude that in order to explain the experimental fusion data at extreme 
subbarrier energies it is not necessary to adopt the hypothesis of an abnormal 
diffuseness of the internuclear potential as advocated in some publications.
Such an assumption would be equivalent, within the folding model, to an unusual 
diffuseness of the target and projectile nuclear matter distributions.
Proton diffusivities, as inferred from electron scattering measurements 
($^{64}$Ni) or from HFB calculations ($^{58}$Ni and $^{100}$Mo) are able to 
reproduce the excitation functions. 
Using a diffuseness 
of the neutron matter distribution in the range of 0.5-0.6 fm, which is a 
reasonable assumption for the moderately neutron rich nuclei considered 
in this paper, is necessary not only to realize a shallow pocket but also 
to obtain a good value for the height of the spherical Coulomb barrier.
With these remarks we do not want to discard the appearance of a large diffuseness 
during the fusion process. Physically we expect an increase in the nuclear skin to 
appear in the neck (overlap) region but not over the entire surface of the reacting 
nuclei.
 
We have tried to advocate in this work that the understanding of the 
experimental data requires more than simply a modification in the 
curvature of the barrier
(as used in the Hill-Wheeler approximation of the Wong formula), 
or the introduction of a simple recipe for the repulsive potential like 
a gaussian. We used the double-folding potential that is based on the Reid 
parametrization of the M3Y interaction, and realistic parameters of the 
proton and neutron distributions of both target and projectile.
We have supplemented this potential  with a repulsive potential that takes 
into account the incompressibility of the nuclear matter.
 
We also conclude that it is necessary to define the fusion in terms of 
in-going-wave boundary conditions. When these consitions are imposed at 
the minimum of the potential pocket, it is possible to reproduce the steep 
energy dependence of the fusion data at extreme subbarrier energies.
Simulating the fusion by the absorption in an imaginary potential, 
on the other hand, does not allow us to reproduce the steep falloff 
of data at the lowest energies.

In the two cases that recently were measured at ATLAS, namely, 
$^{64}$Ni+$^{64}$Ni and $^{64}$Ni+$^{100}$Mo, 
our calculations show a single broad maximum in the $S$-factor at the lowest
measured energies in agreement with the measurements. 
In contrast, our calculations for $^{58}$Ni+$^{58}$Ni show a double-peaked 
structure of the $S$-factor. 
We find that the predicted shape of the low-energy $S$-factor is very sensitive
to the couplings to the two-phonon octupole states. Thus if we ignore these
couplings, the $S$-factor develops a double-peaked structure at low energy, 
whereas a strong coupling to the these states (as in a harmonic vibration) 
tends to produce a single, broad low-energy peak.
It is therefore important to test the predicted $S$-factor experimentally 
at lower energies, in particular in the case of $^{58}$Ni+$^{58}$Ni. 

We noticed for all three fusion reactions studied in this paper a nice 
correlation between the minimum of the potential pocket $V_{\rm min}$ and the 
experimentally extracted reference energy $E_s$, where the $S$-factor reaches
a maximum.
A  systematic study of fusion reactions over a wider range of values of the 
parameter $Z_1Z_2\sqrt{\mu}$ is  necessary to confirm this apparent correlation,
and the conjecture of a repulsive core  for overlapping configurations.
We also mentioned another possible signature of a shallow pocket,
which is the narrowing of the spin distribution at energies below $E_s$. 
This is what our C.~C. calculations predict and it can be tested by 
measurements of the $\gamma$-ray multiplicity emitted from the compound
nucleus.

Before submitting this paper further evidence on the hindrance in the
sub-barrier fusion of $^{48}$Ca+$^{96}$Zr was reported \cite{stef06}. 
This conclusion resulted from the analysis of the 
logarithmic derivative which exhibits a steep increase at the lowest 
measured cross sections.

\section*{Acknowledgements}
One of the authors (\c S.M.) is grateful to the Fulbright Commission 
for financial support and for the hospitality of the Physics Division at 
Argonne National Laboratory where this work was completed. 
H.E. acknowledge the support of the U.S. Department of Energy, 
Office of Nuclear Physics, under Contract No. 
DE-AC02-06CH11357.
We are also grateful to C. L. Jiang, B. B. Back, R. V. F. Janssens 
and K. E. Rehm for constructive comments on this article.

\end{document}